\documentclass[aps,prb,reprint,floatfix,showpacs,superscriptaddress,fleqn]{revtex4-1}
\usepackage{amssymb}

\usepackage{graphicx}% Needed to include png/jpg/pdf figure files
\usepackage{amsmath}
\usepackage{bm}
\usepackage{color}
\usepackage{siunitx}

\usepackage[firstpage]{draftwatermark}%only first page
\usepackage[breaklinks]{hyperref}%hyperef package does internal and external linking (labels, urls)
\usepackage[all]{hypcap}%addition to hyperref, it ensures figure is correctly displayed (and not only the caption) when link is clicked

\hypersetup{
    plainpages=false,
    bookmarks=false,         % show bookmarks bar?
    unicode=false,          % non-Latin characters in Acrobats bookmarks
    pdfborder={0 0 0}
    pdftoolbar=true,        % show  toolbar?
    pdfmenubar=true,        % show  menu?
    pdffitwindow=false,     % window fit to page when opened
    pdfstartview={FitH},    % fits the width of the page to the window
    pdftitle={BP_v2},    % title
    pdfauthor={Vache Folle},     % author
    pdfproducer={LNCMI-CNRS-UGA-UPS-INSA}, % producer of the document
    pdfkeywords={High} {Magnetic} {Fields}, % list of keywords
    pdfnewwindow=true,      % links in new window
    linktoc=section,
    colorlinks=true,       % false: boxed links; true: colored links
    linkcolor=blue,          % color of internal links
    citecolor=red,        % color of links to bibliography
    filecolor=magenta,      % color of file links
    urlcolor=blue           % color of external links
}

\begin{document}

\title{Onset of exciton-exciton annihilation in single layer black phosphorus}

\author{A.\ Surrente}
\affiliation{Laboratoire National des Champs Magn\'etiques Intenses,
UPR 3228, CNRS-UGA-UPS-INSA, Grenoble and Toulouse, France}

\author{A.\ A.\ Mitioglu}\altaffiliation[Present address: ]{High Field Magnet Laboratory (HFML-EMFL), Institute for Molecules and Materials, Radboud University, Toernooiveld 7, 6525 ED Nijmegen, The Netherlands}
\affiliation{Laboratoire National des Champs Magn\'etiques Intenses,
UPR 3228, CNRS-UGA-UPS-INSA, Grenoble and Toulouse,
France}\affiliation{Institute of Applied Physics, Academiei Str.\ 5,
Chisinau, MD-2028, Republic of Moldova}

\author{K.\ Galkowski}
\affiliation{Laboratoire National des Champs Magn\'etiques Intenses,
UPR 3228, CNRS-UGA-UPS-INSA, Grenoble and Toulouse, France}

\author{L.\ Klopotowski}
\affiliation{Institute of Physics, Polish Academy of Sciences, al. Lotnik\'{o}w 32/46, 02-668 Warsaw, Poland}

\author{W.\ Tabis}
\affiliation{Laboratoire National des Champs Magn\'etiques Intenses,
UPR 3228, CNRS-UGA-UPS-INSA, Grenoble and Toulouse, France}
\affiliation{AGH University of Science and Technology, Faculty of Physics and Applied Computer Science, Al.\ Mickiewicza 30, 30-059 Krakow, Poland}

\author{B.\ Vignolle}
\affiliation{Laboratoire National des Champs Magn\'etiques Intenses,
UPR 3228, CNRS-UGA-UPS-INSA, Grenoble and Toulouse, France}

\author{D.\ K.\ Maude}
\affiliation{Laboratoire National des Champs Magn\'etiques Intenses,
UPR 3228, CNRS-UGA-UPS-INSA, Grenoble and Toulouse, France}

\author{P.\ Plochocka}\email{paulina.plochocka@lncmi.cnrs.fr}
\affiliation{Laboratoire National des Champs Magn\'etiques Intenses,
UPR 3228, CNRS-UGA-UPS-INSA, Grenoble and Toulouse, France}

%\SetWatermarkText{FIRST DRAFT PLEASE DO NOT DISTRIBUTE}%Text for watermark
%\SetWatermarkLightness{0.9}%How light/dark
%\SetWatermarkScale{0.25}%Font scaling

\date{\today}

\begin{abstract}
The exciton dynamics in monolayer black phosphorus is investigated
over a very wide range of photoexcited exciton densities using time
resolved photoluminescence. At low excitation densities, the exciton
dynamics is successfully described in terms of a double exponential
decay. With increasing exciton population, a fast, non-exponential
component develops as exciton-exciton annihilation takes over as the
dominant recombination mechanism under high excitation conditions.
Our results identify an upper limit for the injection density, after
which exciton-exciton annihilation reduces the quantum yield, which
will significantly impact the performance of light emitting devices
based on single layer black phosphorus.
\end{abstract}

\maketitle 

\section{Introduction}\label{sec:intro}
The reduced dimensionality of atomically thin two-dimensional (2D)
semiconductors enhances many-body interactions, which influence
strongly the carrier dynamics at high carrier concentrations. The
large exciton binding energy in these layered materials
\cite{he2014tightly,hill2015observation,Tran14,Castellanos14}
facilitates the observation of such many-body interactions in the
exciton physics. In particular, exciton-exciton annihilation is a
scattering mechanism in which one exciton recombines
non-radiatively, transferring its energy and momentum to another
exciton, which subsequently relaxes to lower energy states, losing
the energy initially gained, through electron-phonon interactions.
This process, which can be viewed as the exciton counterpart of
Auger recombination for free carriers, has been demonstrated to be a
very efficient non radiative recombination channel (particularly at
high injection levels) in low dimensional systems such as carbon
nanotubes \cite{Lueer09,Wang04,Huang06,Ma05}, graphene nanoribbons
\cite{soavi2016exciton}, colloidal quantum dots and nanorods
\cite{Klimov00,Htoon03}, organic
semiconductors~\cite{Shaw08,akselrod2010exciton}, and monolayer
transition metals dichalcogenides (TMDs)
\cite{Kumar14,Sun14,yu2016fundamental,Yuan15,Mouri14}. Other than the
understanding of fundamental exciton dynamics, the interest of
exciton-exciton annihilation resides in its significant
technological impact, since it provides an upper limit for the
carrier density under different excitation conditions.
Exciton-exciton annihilation is therefore expected to have a
significant influence on the efficiency of light emitting devices,
such as semiconductor optical amplifiers and semiconductor lasers.
It is also of interest in view of the production of highly efficient
photovoltaic devices based on 2D materials. One of the factors
limiting the efficiency of conventional solar cells is represented
by the excess kinetic energy, dissipated as heat, of the hot
carriers generated by the absorption of a photon with an energy well
above the band gap. Efficient exciton-exciton annihilation might
also imply an efficient multiple exciton generation
\cite{Schaller04,padilha2013aspect,davis2015multiple} (\emph{i.e.}
time-reversed exciton-exciton annihilation) which could potentially
lead to the harvesting of this excess energy, significantly reducing
the influence of this loss channel.

Black phosphorus is a recent member of the rapidly expanding family
of atomically thin 2D semiconductors obtained by mechanical
exfoliation \cite{ling2015,Gomez15}. Its electronic properties
combine the advantages of graphene in terms of carrier mobility with
the direct band gap observed in monolayer TMDs. This results in an
on/off ratio for black phosphorus field effect transistors of the
order of 10$^{5}$ \cite{ling2015,li14,Liu14}. The band gap of black
phosphorus can be tuned from its bulk value (\SI{0.3}{\eV}) to the
visible part of the spectrum by decreasing the number of layers
\cite{Tran14,ling2015}. As in TMDs, the strong carrier confinement
and the decreased Coulomb screening due to the reduced
dimensionality lead to an optical response of monolayer black
phosphorus dominated by excitonic effects (the theoretical exciton
binding energy is \SI{0.8}{\eV} in vacuum \cite{Tran14} and
\SI{0.38}{\eV} on a Si/SiO$_{2}$ substrate \cite{Castellanos14}).
The neutral exciton was reported to exhibit a photoluminescence (PL)
emission energy of $\lesssim\SI{2}{\eV}$ in monolayer samples
\cite{Liu14,wang2015highly,yang2015optical,surrente16}. Such
properties make atomically thin black phosphorus an extremely
promising material in electronics, for example to manufacture field
effect transistors,\cite{xia14,ling2015,li14,Liu14} and in
optoelectronics with potential applications as
photodetectors\cite{Buscema15}, or photon
polarizers\cite{Tran14,Low14,xia14,Qiao14,ling2015}. The
responsivity of black phosphorus based photodetectors has been shown
to decrease with increasing excitation power \cite{Buscema15}.
Although this effect has been attributed to a decrease of the
available carriers created under high excitation conditions
\cite{Buscema15}, the specific mechanism (saturation of trap states
or Auger-like recombination) has yet to be unequivocally identified.
As atomically thin black phosphorus has been successfully isolated
only recently, very little is currently known about the carrier
dynamics. Using pump-probe measurements, the evolution of exciton
dynamics has been investigated, focusing on the in plane anisotropy
\cite{Ge15,Suess15,He15}. Most experiments have been performed at
room temperature, and on thick samples \cite{Ge15}, where the many
body effects are detrimentally weakened when compared to what is
expected at low temperature on single atomic layers. One study
reports the temporal evolution of the PL of single layer black
phosphorus, investigated over a limited range of excitation powers,
always in the low excitation regime, with a single exponential PL
decay \cite{yang2015optical}.

In this paper, we present power dependent time resolved micro-PL
(\si{\micro}PL) performed at low temperature. The pump fluence
employed spans over almost four orders of magnitude, thereby
providing access to the exciton dynamics in very different
excitation regimes. We show that at low excitation densities the
exciton dynamics in monolayer black phosphorus can be described
using a double exponential decay. The first, shorter decay time, of
the order of tens of picoseconds, is attributed to defect-mediated
processes \cite{shi2013exciton,wang2014ultrafast}. The longer decay
time, of the order of a few hundreds of picoseconds, is signature of
the excitonic radiative recombination \cite{shi2013exciton}. In the
high fluence regime, both decay times decrease, pointing to the
onset of many-body processes. Finally, at a yet higher exciton
population, a fast, non-exponential decay component develops. Under
this condition, the observed temporal dependence is best described
by a bimolecular model involving exciton-exciton annihilation.

\section{Experimental details}\label{sec:ExperimentalDetails}
The bulk black phosphorus has been purchased from purchased from Smart Elements (99.998\% nominal purity). The mechanical exfoliation of the flakes was performed in a protected atmosphere (glove box filled with Ar gas, $<\SI{1}{ppm}$ O$_2$, $<\SI{1}{ppm}$ H$_2$O). The flakes were subsequently transferred onto a Si substrate capped with a \SI{300}{\nano\m} thick layer of thermal SiO$_2$. These samples were stored in vials, always in the glove box, and transferred to the He-flow cryostat used for the optical characterization, always under an Ar atmosphere. After mounting the samples, the cryostat was rapidly pumped to a pressure below $\SI{1e-4}{\milli\bar}$, minimizing the exposure of the black phosphorus to air. The single layer character of the samples was verified by Raman measurements performed in situ on the same flake where the PL was investigated \cite{surrente16}.

The steady-state \si{\micro}PL signal was excited with a \SI{532}{\nano\m} laser and the spectra were recorded using a
spectrometer equipped with a liquid nitrogen cooled CCD camera. The time-resolved \si{\micro}PL signal was excited with
the frequency doubled output of a tuneable optical parametric oscillator (OPO), synchronously pumped with a mode-locked
Ti:Sapphire laser. The typical temporal pulse width was $\simeq \SI{300}{\femto\s}$, with a repetition rate of
\SI{80}{\mega\Hz}. The time resolved PL signal was dispersed by an imagining spectrometer and detected using a
synchroscan streak camera with the temporal resolution set to $\sim\SI{10}{\pico\s}$. All the measurements presented
here have been performed at \SI{4.2}{\K}.

\section{Results and discussion}\label{sec:results}
\begin{figure}[htb]
    \begin{center}
        \includegraphics[width=1.0\linewidth]{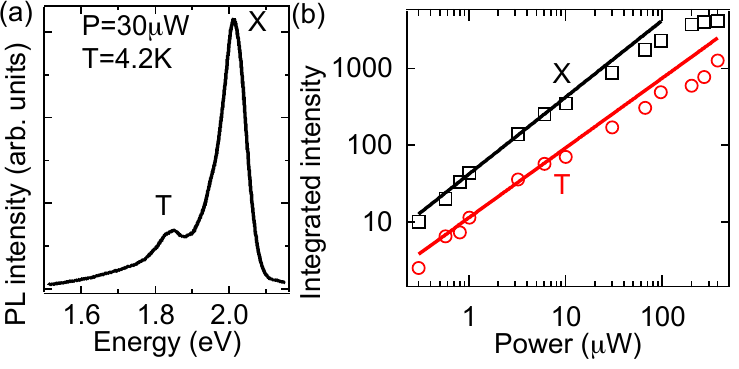}
    \end{center}
\caption{(a) Time integrated \si{\micro}PL spectrum of monolayer black phosphorus. (b) Integrated intensity of the
neutral (X) and charged (T) exciton versus power. The solid lines represent fits to a power law.}
\label{fig:Pdependence}
\end{figure}

The static excitonic response of the investigated flake was probed
by time-integrated \si{\micro}PL spectroscopy. A representative, low
temperature \si{\micro}PL spectrum of the monolayer black phosphorus
flake is shown in Fig.\ \ref{fig:Pdependence}(a) (excitation power
$P=\SI{30}{\micro\W}$). The spectrum consists of two distinct peaks,
resulting from the recombination of the neutral (X) and charged (T)
exciton.

In Fig.\,\ref{fig:Pdependence}(b), we show the integrated intensity
of the peaks versus excitation power. We fitted the integrated
intensity $I$ to the power law $I \propto P^n$, obtaining $n=1.00
\pm 0.02$ for the neutral exciton and $n=0.91 \pm 0.05$ for the
charged exciton for $P\leq\SI{10}{\micro\W}$. This confirms that in
both cases the recombination process involves a single electron-hole
pair \cite{surrente16}. At higher excitation power, the integrated
intensity of this flake shows a sublinear increase as a function of
the excitation power, with a clear saturation of the intensity of
the neutral exciton for $P\geq\SI{200}{\micro\W}$. Considering that
the band gap of black phosphorus (and hence its emission energy) is
known to blue shift with increasing temperature
\cite{warschauer1963electrical,baba1991photoconduction,surrente16},
we can rule out laser-induced heating effects even at the highest
excitation intensity, as no shift of the PL emission energy is
observed for this sample. This, along with the constant line shape
\cite{surrente16} (pointing to the absence of biexciton
recombination), the saturation of the PL intensity at high
excitation power and the 2D excitonic character of the emission of black phosphorus, suggests that other non-linear processes, namely
exciton-exciton annihilation, play an important role \cite{Yuan15,Kumar14,akselrod2010exciton,yu2016fundamental,Mouri14,Sun14}.

\begin{figure}[htb]
    \begin{center}
        \includegraphics[width=1.0\linewidth]{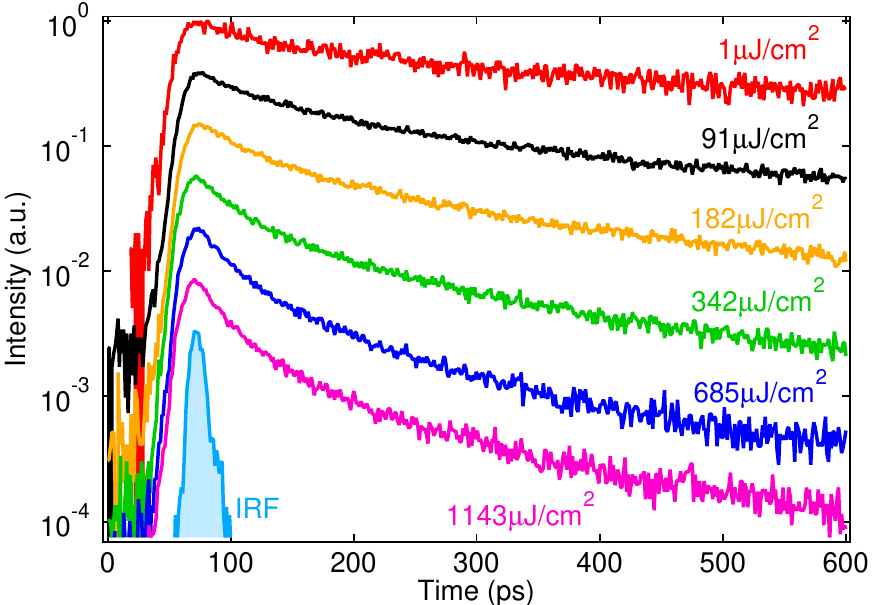}
    \end{center}
\caption{Time decay of the neutral excitonic transition at various pump fluences. The instrument response
function (IRF) is shown for reference. The curves are vertically offset for clarity.} \label{fig:TRpowerDependenceBP}
\end{figure}

To elucidate the origin of the observed non-linear behavior of the
black phosphorus PL, we performed systematic measurements of
time-resolved \si{\micro}PL at a varying excitation density (see
Fig.\,\ref{fig:TRpowerDependenceBP}). For reference, in
Fig.\,\ref{fig:TRpowerDependenceBP} we also show the temporal
evolution of the excitation pulse, which represents the lower limit
of the temporal resolution of our detection system (instrument
response function, IRF, with a full width at half maximum of
$\sim\SI{10}{\pico\s}$, as extracted from a Gaussian fit). We note
that even at the lowest pump fluence, the time decay of the neutral
exciton is characterized by two different slopes, indicating the
presence of at least two decay components. Moreover, with increasing
fluence, the fast decay component becomes more significant, as
attested by the prominent peak of the time-resolved traces at short
delay times, whereas the slower decay components tends to decrease,
with a corresponding smaller slope observed at larger delay times at
high power.

\begin{figure}[htb]
    \begin{center}
        \includegraphics[width=1.0\linewidth]{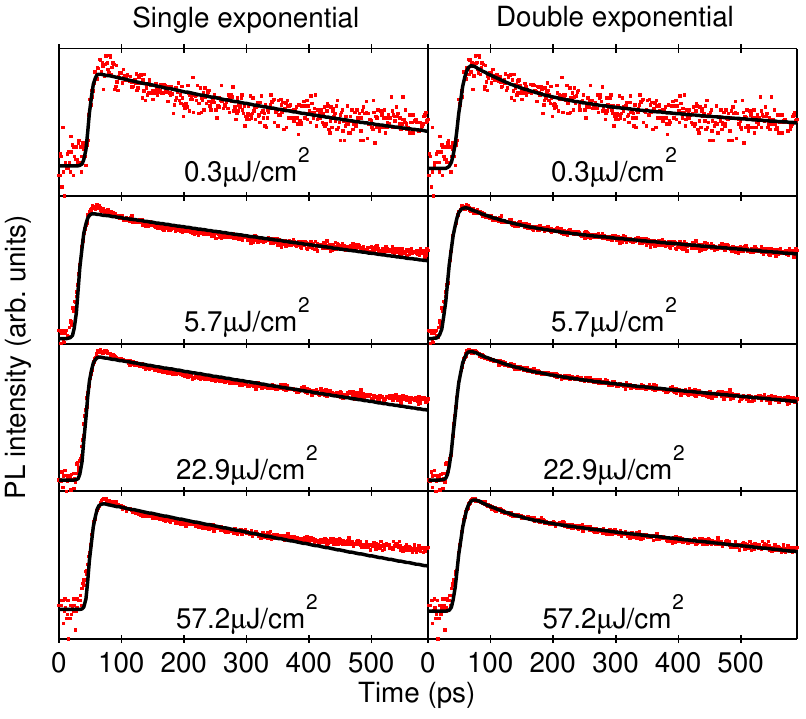}
    \end{center}
\caption{Time resolved \si{\micro}PL measurements of neutral exciton (dotted curves) and fits to single exponential and double exponential models (solid lines). The pump fluences are indicated in each panel.}
\label{fig:SingleDoubleExp}
\end{figure}

\begin{figure}[htb]
    \begin{center}
        \includegraphics[width=1.0\linewidth]{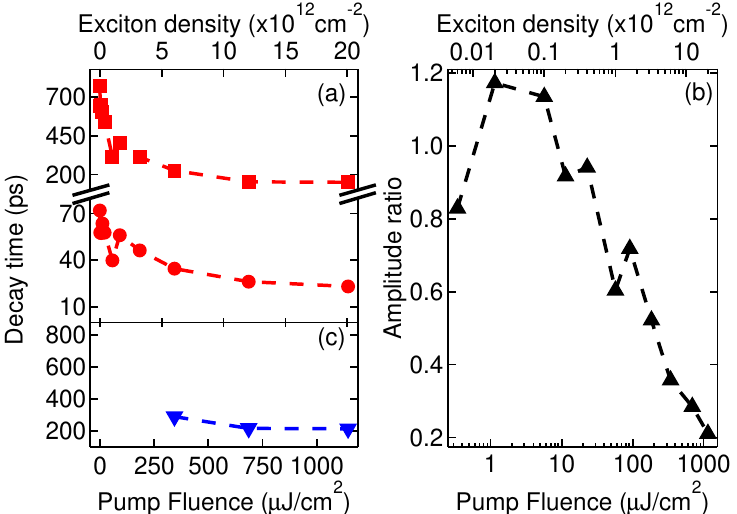}
    \end{center}
\caption{(a) Decay times and (b) ratio between the two exponential components versus pump fluence and initial exciton density, double exponential model. (c) Decay time as a function of pump fluence and initial exciton density, bimolecular model. The dashed lines are a guide to the eye.}
\label{fig:TauGammaVsPower}
\end{figure}

To verify whether the time decay at low pump fluence (or
equivalently at low exciton density) could be described in terms of
only radiative recombination, we have fitted the PL intensity $I(t)$
to a single exponential decay convoluted with a Gaussian IRF,
$I(t)=\int_{-\infty}^{t}
\text{IRF}(t')\text{e}^{-(t-t')/\tau}\text{d}t'$. In the left column of Fig.\ \ref{fig:SingleDoubleExp} we present the results of these fits. Except for the lowest fluence, the single exponential model is unable to fit satisfactorily the data, showing large difference with respected to the measured decay especially at very short and at long $t$. This demonstrates that a second relaxation mechanism (other than electron-hole recombination) has to play a role in the investigated sample. A wide range of defects or impurities are expected to
influence the electronic properties of mechanically exfoliated black
phosphorus monolayers \cite{yuan2015transport}. Such defects could
induce trap states for the excitons, which are periodically
populated and depopulated. The combination of this relaxation path
with the electron-hole recombination can be accounted for by a
double exponential decay, in which the PL intensity is described by
$I(t)=\int_{-\infty}^{t}
\text{IRF}(t')\left[\text{e}^{-(t-t')/\tau_{\text{d}}}+ a
\text{e}^{-(t-t')/\tau}\right]\text{d}t'$, where $a$ denotes the
ratio between the two exponential components, $\tau_{\text{d}}$ and
$\tau$ are decay times due to the defect related path and radiative
recombination respectively. Using this model to describe the data
leads to a considerably improved quality of the fits, at least at
low and intermediate pump fluence. This is demonstrated by the fits presented in the right column of Fig.\ \ref{fig:SingleDoubleExp}, which capture the time decay of the PL signal significantly better, suggesting that defect-mediated trapping relaxation plays an important role in the mechanically exfoliated black phosphorus monolayers. We attribute the fast decay
to trapping or defect-assisted recombination mechanisms, as
previously reported in TMDs \cite{shi2013exciton,wang2014ultrafast},
while the slow decay reflects the exciton radiative recombination
\cite{shi2013exciton}. Our results at low fluence are not
surprising, since similar, multiexponential decays have already been
reported for black phosphorus \cite{Ge15,Suess15,He15}. Fitting with
the double exponential model all the data measured at a different
excitation power, gives values of $\tau_{\text{d}}$ of tens of
picoseconds and values of $\tau$ of hundreds of picoseconds [see
Fig.\ \ref{fig:TauGammaVsPower}(a) for the exact values]. 

\begin{figure}[htb]
    \begin{center}
        \includegraphics[width=1.0\linewidth]{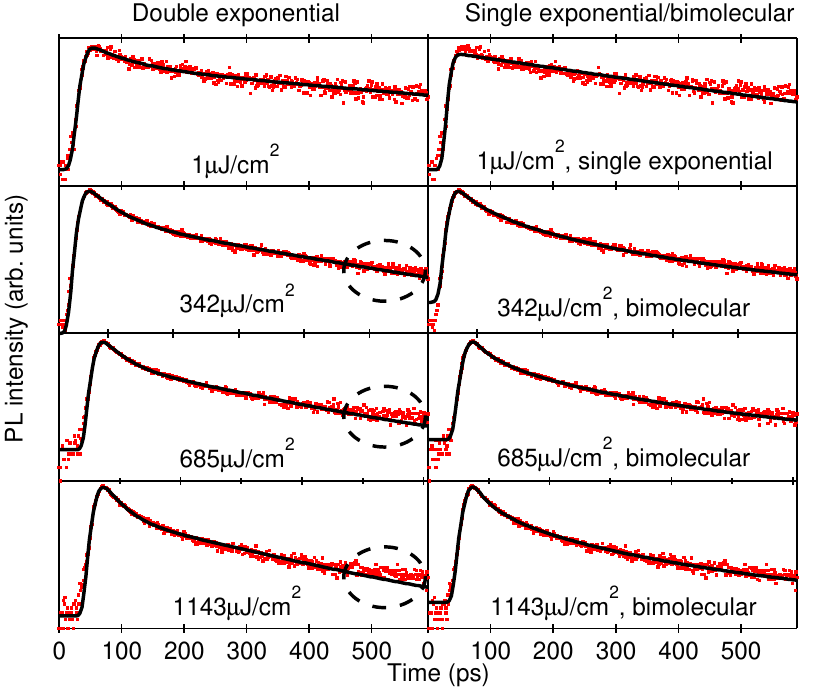}
    \end{center}
\caption{Time decay of neutral exciton (dotted curves) and fits to single exponential, double exponential and
bimolecular models (solid lines). The dashed ellipses highlight the delay times at which the double exponential mode is
unable to fit the data.} \label{fig:DoubleExpVsBimolecular}
\end{figure}

The time resolved \si{\micro}PL measured at high fluence is shown Fig.\ \ref{fig:DoubleExpVsBimolecular}. Obtaining
the best fit with a double exponential model (see left column of Fig.\ \ref{fig:DoubleExpVsBimolecular}) requires a varying ratio $a$ between components with
increasing excitation, as summarized in Fig.\
\ref{fig:TauGammaVsPower}(b). We note that if the exciton dynamics
was solely dominated by the two mechanisms mentioned above, the
saturation of the exciton traps at very high fluence would lead to
an increasing weight of the radiative recombination component $a$
\cite{Sun14}. Here, at the lowest pump fluences $a$ increases
slightly and then remains essentially constant up to
$\sim\SI{20}{\micro\J/\centi\m^2}$, indicating that the trap states
are being gradually filled, although probably not up to the full
saturation. At higher pump fluence the opposite trend is
experimentally observed, with the amplitude ratio decreasing
significantly with increasing fluence [see
Fig.\,\ref{fig:TauGammaVsPower}(b)]. This, along with variation of
the extracted decay times $\tau$ and $\tau_{\text{d}}$ as a function
of the pump fluence [see Fig.\,\ref{fig:TauGammaVsPower}(a)],
suggests the occurrence of non-linear processes coming into play
when the exciton population is sufficiently large. This is
corroborated by the failure of the double exponential model to fit
the exciton dynamics measured at pump fluence larger than
$\geq\SI{342}{\micro\J/\centi\m^2}$, as can be seen for example in
the fit of the high fluence time-resolved \si{\micro}PL displayed in
the left column of Fig.\ \ref{fig:DoubleExpVsBimolecular} (note in
particular the discrepancy occurring at large delay times). Such an
excitation level corresponds to the excitation regime at which the
integrated intensity saturates, as shown in
Fig.\,\ref{fig:Pdependence}(b). The observation of a sublinear dependence of the integrated intensity on the excitation power and of the appearance of fast, non-exponential exciton dynamics at the same excitation level has already been interpreted by referring to exciton-exciton annihilation processes \cite{Yuan15}.

We now focus on the exciton dynamics in the non-linear regime. At high pump fluence, the fast decay component becomes
more prominent, as evidenced by the faster initial decay (see Fig.\,\ref{fig:TRpowerDependenceBP}). We rule out the
contribution related to the formation of biexcitonic complexes, which might possibly be responsible for an increasing
fast decay component, as no additional low energy peaks corresponding to biexcitonic recombination are observed at high
power \cite{surrente16}. Instead, we propose exciton-exciton annihilation, already observed in TMDs
\cite{Kumar14,Sun14,yu2016fundamental,Yuan15,Mouri14}, as the recombination mechanism responsible for the observed non-linear
exciton dynamics. In this framework, the rate equation for the exciton population $n(t)$ can be written:
\begin{equation*}
\frac{\partial n}{\partial t} = -\frac{n}{\tau} - \gamma n^2, \label{eq:RateEquation}
\end{equation*}
where $\tau$ is the exciton lifetime, and $\gamma$ is the
annihilation rate. At high fluences, the quadratic (bimolecular)
term leads to a rapid decay which dominates over the defect channel
which can then be neglected. We would like to stress that the
bimolecular term is too small to explain the fast decay at low
fluences, for which the double exponential model, which includes the
defect channel, is required.

Assuming that $\gamma$ is independent of time, and in the regime of very short excitation pulse
($\tau_{\text{pulse}}\ll \tau$), the PL temporal dependence $I_{\text{bm}}(t)$ in the bimolecular model can be written as
\cite{Shaw08, Yuan15, akselrod2010exciton}
\begin{equation*}
I_{\text{bm}}(t)=\frac{I_{\text{in}}\text{e}^{-t/\tau}}{1+\gamma\tau n(0)\left(1-\text{e}^{-t/\tau}\right)},
\label{eq:n_t}
\end{equation*}
where $I_{\text{in}}$ denotes the pump intensity and $n(0)$
represents the initial exciton density, estimated
\cite{jordan1999carrier} by taking into account the Gaussian profile
of the excitation spot, its size ($\sim\SI{1}{\micro\m}$), the excitation wavelength (\SI{526}{\nano\m}) and the
absorbance of monolayer black phosphorus of 2.8\% per layer
\cite{Castellanos14}. By convoluting $I_{\text{bm}}(t)$ with the
IRF, we have extracted the exciton lifetimes and the annihilation
rates for pump fluences larger than \SI{342}{\micro\J/cm^2}. We
found a decay time $\tau\lesssim\SI{280}{\pico\s}$ [see
Fig.\,\ref{fig:TauGammaVsPower}(c) for the detailed fluence
dependence], similar to the $\tau$ obtained at low pump fluence in
the double exponential model. 
%At high pump fluence,
%($>\SI{300}{\micro\J/\centi\m^2}$), $\tau$ is found to decrease with
%increasing pump fluence (which points to more pronounced
%exciton-exciton annihilation effects), before saturating at
%$\tau\sim\SI{260}{\pico\s}$. 
All the high fluence decays could be
fitted using an almost constant value of  annihilation rate
$\gamma=5.0\pm\SI{0.2e-3}{\centi\m^2/\s}$. Therefore, if we restrict
our analysis to the highest excitation conditions, we can
successfully fit the exciton dynamics with a single set of
parameters, suggesting that exciton-exciton annihilation becomes the
dominant recombination mechanism in the investigated sample at an
exciton concentration of $n(0)\sim\SI{6.1e12}{\centi\m^{-2}}$. The validity of the bimolecular model at very high pump fluence is further corroborated by the comparison of $\tau_{\text{d}}$ with the effective, exciton-density dependent fast time constant resulting from the bimolecular model $(\gamma n)^{-1}$. For fluences larger than \SI{340}{\micro\J/\centi\m^2}, our fitting parameters show that $(\gamma n)^{-1}<\tau_{\text{d}}$ (at the highest exciton densities by more than a factor of 2), demonstrating that at these excitation conditions the fast exciton dynamics is dominated by non-linear processes, and allowing us to neglect the influence of defect-mediated recombination. The
annihilation rate is related to the exciton diffusion constant $D$
and to the maximum distance $r_{\text{a}}$ at which the annihilation
process occurs  through $\gamma=4 \pi D r_{\text{a}}$
\cite{shaw2010exciton}. The substrate could affect the measured
annihilation rate either by reducing the binding energy
\cite{Castellanos14}, which might lead to a decrease of
$r_{\text{a}}$, or through a reduced mobility \cite{ma2014charge},
which can influence the exciton diffusion constant. Moreover, the
presence of defects in the mechanically exfoliated flakes introduces
an additional recombination pathway in competition with the
exciton-exciton annihilation, partially smearing its effects
\cite{yu2016fundamental}. This suggests that in the case of a better crystalline
quality of monolayer black phosphorus, exciton-exciton annihilation
could occur at lower exciton densities.

The recently reported annihilation rates of monolayer TMDs range
from $\sim0.04-\SI{0.1}{\centi\m^2/\s}$ for MoS$_2$
\cite{Sun14,yu2016fundamental} to $\sim0.3-\SI{0.4}{\centi\m^2/\s}$ for WSe$_2$
\cite{Mouri14}, WS$_2$ \cite{Yuan15,yu2016fundamental}, and MoSe$_2$
\cite{Kumar14}, with a typical exciton density of the order of
$\SI{10e10}{\centi\m^{-2}}$
\cite{Sun14,yu2016fundamental,Mouri14,Yuan15,Kumar14}. The lower annihilation
rate of monolayer black phosphorus of
$5.0\pm\SI{0.2e-3}{\centi\m^2/\s}$ obtained from our measurements is
thus consistent with the larger exciton density that has to be
injected to observe a significant exciton-exciton annihilation.

\section{Conclusions}\label{sec:conclusions}
In summary, we have presented a detailed study of the power
dependence of the neutral exciton dynamics in monolayer black
phosphorus investigated using time-resolved \si{\micro}PL
measurements. Varying the pump fluence over almost four orders of
magnitudes allowed us to probe the exciton dynamics at strongly
different photoexcited exciton densities. At low pump fluence, the
measured PL decay can be satisfactorily fitted with a double
exponential model. In this framework, the fast decay is assigned to
defect-mediated charging and decharging mechanisms, while the
long-lived component is attributed to the radiative recombination of
the excitons. As the pump fluence was increased, the ratio between
the two decay components varies, with the fast decay component
taking over in the high fluence regime, which is the completely
inconsistent with the expected behavior in the case of a saturation
of the trapping defects. The observation of this fast,
non-exponential decay component can be explained in terms of
exciton-exciton annihilation. Our results allow us to identify an
upper limit for the density of injected carriers above which
exciton-exciton annihilation regime completely dominates. The
threshold exciton density is $\sim\SI{6.1e12}{\centi\m^{-2}}$. Above
this density a single set of fitting parameters can be used to fit
the PL decay. The demonstrated domination of exciton-exciton
annihilation has important implications for the optimization of the
quantum yield of advanced light emitting devices and for the
responsivity of photodetectors based on monolayer black phosphorus.

\begin{acknowledgments}
The authors gratefully acknowledge Geert Rikken for providing the bulk black phosphorus. AAM acknowledges financial support from the French
foreign ministry. This work was partially supported by ANR JCJC project milliPICS, the R\'egion Midi-Pyr\'en\'ees under
contract MESR 1305303, STCU project 5809 and the BLAPHENE project under IDEX program Emergence.
\end{acknowledgments}

\bibliography{BlackPDynamicsBib}

%Word count: main text 2792. Captions 149. Figures 198+95+125+99=517. Total 3458
\end{document}